# Anisotropy in the thermal hysteresis of resistivity and charge density wave nature of single crystal SrFeO$_{3-\delta}$: X-ray absorption and photoemission studies

S. H. Hsieh,[1, §] R. S. Solanki,[1, §, †] Y. F. Wang,[1] Y. C. Shao,[1] S. H. Lee,[1] C. H. Yao,[1] C. H. Du,[1] H. T. Wang,[2] J. W. Chiou,[3] Y. Y. Chin,[4] H. M. Tsai,[4] J.-L. Chen,[4] C.W. Pao,[4] C.-M.Cheng,[4] W.-C. Chen,[4] H. J. Lin,[4] J. F. Lee,[4] F. C. Chou,[5] W. F. Pong[1,*]

[1] Department of Physics, Tamkang University, Tamsui 251, Taiwan
[2] Department of Physics, National Tsinghua University, Hsinchu 300, Taiwan
[3] Department of Applied Physics, National University of Kaohsiung, Kaohsiung 811, Taiwan
[4] National Synchrotron Radiation Research Center, Hsinchu 300, Taiwan
[5] Center for Condensed Matter Sciences, National Taiwan University, Taipei 106, Taiwan

The local electronic and atomic structures of the high-quality single crystal of SrFeO$_{3-\delta}$ ($\delta\sim0.19$) were studied using temperature-dependent x-ray absorption and valence-band photoemission spectroscopy (VB-PES) to investigate the origin of anisotropic resistivity in the **ab**-plane and along the **c**-axis close to the region of thermal hysteresis (near temperature for susceptibility maximum, T$_m\sim$78 K). All experiments herein were conducted during warming and cooling processes. The Fe $L_{3,2}$-edge X-ray linear dichrois results show that during cooling from room temperature to below the transition temperature, the unoccupied Fe 3d $e_g$ states remain in persistently out-of-plane $3d_{3z^2-r^2}$ orbitals. In contrast, in the warming process below the transition temperature, they change from $3d_{3z^2-r^2}$ to in-plane $3d_{x^2-y^2}$ orbitals. The nearest-neighbor (NN) Fe-O bond lengths also exhibit anisotropic behavior in the **ab**-plane and along the **c**-axis below T$_m$. The anisotropic NN Fe-O bond lengths and Debye-Waller factors stabilize the in-plane Fe $3d_{x^2-y^2}$ and out-of-plane $3d_{3z^2-r^2}$ orbitals during warming and cooling, respectively. Additionally, a VB-PES study further confirms that a relative band gap opens at low temperature in both the **ab**-plane and along the **c**-axis, providing the clear evidence of the charge-density-wave nature of SrFeO$_{3-\delta}$ ($\delta\sim 0.19$) single crystal.

Recently, $SrFeO_{3-\delta}$ based colossal magnetoresistance (CMR) materials have attracted much interest owing to their potential applications in the next generation of magnetic data storage read heads, and because of the research goal of elucidating the microscopic origin of CMR.[1-4] Extensive research on these CMR materials and associated electron structures that has been performed for the last decade is still ongoing.[5-10]

The structural, magnetic and transport properties of $SrFeO_{3-\delta}$ vary dramatically with oxygen content and the valence state of Fe.[7,11-13] The known phases of $SrFeO_{3-\delta}$ include stoichiometric $SrFeO_3$ ($\delta= 0$), which has a cubic perovskite structure with a valence state of $Fe^{4+}$, and oxygen-deficient phases with tetragonal ($\delta= 0.125$), orthorhombic ($\delta= 0.25$) and brownmillerite-type ($\delta= 0.50$) crystal structures.[1,2] Certain amount of oxygen result in the coexistence of these phases. The tetragonal phase comprises $FeO_6$ octahedra, distorted/tilted $FeO_6$ octahedra and square pyramidal $FeO_5$.[2] The orthorhombic phase reportedly includes distorted $FeO_6$ octahedra and square pyramidal $FeO_5$, in which Fe has valence of 3+ and 4+, respectively. The brownmillerite-type phase has tetrahedrally and octahedrally coordinated Fe sites with an Fe valence state of 3+.[2] In these oxygen-deficient $SrFeO_{3-\delta}$ systems, giant negative magnetoresistance has been observed in the tetragonal phase around a coincident charge-spin ordering transition temperature.[1,2]

This transition exhibits thermal hysteresis of resistivity and magnetization, suggesting a first-order phase transition. The thermal hysteresis around the transition temperature (close to the temperature for susceptibility maximum, $T_m$) has been attributed to the coexistence of antiferromagnetic and paramagnetic phases.[5,6] Further, a change in the lattice structure that originates from the charge ordering (CO) transition has been observed in Raman, far-infrared ellipsometric and neutron diffraction studies.[1,2,14] In transition metal oxides, CO-associated lattice distortion, which depends on the electron-lattice coupling strength, can cause orbital ordering (OO) so CO has been found to coexist with OO.[15-18] Bao et al. presented evidence of a CO-induced structural phase transition in single crystalline $(Bi,Ca)MnO_3$.[15] Chi et al. also observed strong enhancement of orthorhombic distortion owing by CO, which allows OO to be established in single crystals of $Pr_{0.5}Ca_{1.5}MnO_4$.[17] Similar observations have been made concerning other CO transitions, specifically, the charge-disproportionation of Fe ion and/or the formation of a charge-density-wave (CDW) in $La_{1-x}Sr_xFeO_3$.[3,19] However, in $(La_{2-2x}Sr_{1+2x})Mn_2O_7$, competing local lattice distortion and the hopping of charge carriers generate comparable amounts of Mn out-of-plane $3d_{3z^2-r^2}$ and in-plane $3d_{x^2-y^2}$ orbitals in the metallic phase.[16] Therefore, the electrical and magnetic properties of $(La_{2-2x}Sr_{1+2x})Mn_2O_7$ depend on not only the out-of-plane but also the in-plane Mn $3d$ states. Clearly, these orbital effects influence the electronic properties of the sample, resulting in peculiar behavior that is associated with the resistivity-related phenomena.[20-22] Chen *et al.* used a theoretical model of iron pnictides to show that unequal $d_{xz}$ and $d_{yz}$ orbital populations can result in anisotropic resistivity.[23] They further noted that such preferable orbital states can be identified by x-ray linear dichroism (XLD).[23] However, the complex interplay between the mechanisms that drive OO and cause these orbital effects remains unresolved.[24-26] As discussed above, the tetragonal phase of $SrFeO_{3-\delta}$ undergoes a first-order spin-charge ordering transition that is accompanied by a change in lattice structure, which exhibits exotic electrical and magnetic properties.[1] Most spectroscopic investigations of $SrFeO_{3-\delta}$-based materials have involved varying the oxygen content and substitution at Sr or Fe sites; very few experimental and theoretical studies that involve the metal-to-insulator (MI)/semiconductor transition in $SrFeO_{3-\delta}$.[3,4,19] Using resonant x-ray scattering (RXS), Martin *et al.* elucidated the CDW nature of the CO transition (MI/semiconductor transition) in La-substituted $SrFeO_3$.[19] However, their studies did not provide evidence of band gap opening in the formation of CDW.[19]

In this context, an investigation of the local electronic and atomic structures of tetragonal $SrFeO_{3-\delta}$ by x-ray polarization (***E***)-dependent absorption and valence-band photoemission spectroscopy (VB-PES), can provide insight into the preferred orbital of Fe 3*d* states, lattice distortion that is related to the anisotropy of resistivity in the thermal hysteresis region, and band gap opening as a function of temperature. Therefore, a detailed study of single crystalline $SrFeO_{3-\delta}$ ($\delta \sim 0.19$) is conducted herein using x-ray diffraction, magnetization, resistivity, x-ray absorption near-edge spectroscopy (XANES), XLD, extended x-ray absorption fine structure (EXAFS) and VB-PES. All experiments were carried out during warming and cooling runs with normal incidence (***E***//***ab***-plane, $\theta = 0^0$) and incidence at a glancing angle (near ***E***//***c***-axis, $\theta = 70^0$).

**Results and discussion**

Figure 1(a) presents Lebail fits[27-29] of the X-ray powder diffraction (XRPD) of $SrFeO_{3-\delta}$ sample at room temperature based on the tetragonal *I4/mmm* space group model.[12] The refined lattice parameters are ***a*** = ***b*** = 10.9320(5) Å and ***c*** = 7.7012(5) Å. The insets in figure 1(a) magnify some of the Bragg reflections. The singlet nature of the $(111)_{pc}$ peak and the doublet nature of the $(200)_{pc}$ and $(220)_{pc}$ (pc= pseudocubic) peaks are characteristic of the tetragonal phase. Figure 1(b) schematically depicts the crystal structure of the tetragonal $SrFeO_{3-\delta}$, which includes Fe sites of three types pyramidal, [Fe(1), figure 1(c)], distorted/tilted octahedral, [Fe(2), figure 1(d)] and octahedral [Fe(3), figure 1(e)] with valence states of $Fe^{4+}$, $Fe^{3.5+}$ and $Fe^{4+}$, respectively on the basis of the Mössbauer spectroscopy results that were obtained by Lebon *et al*.[1]

and the bond valence calculations that were made by Hodges *et al.*[12]

Figure 2 plots the temperature-dependence of resistivity in the *ab*-plane and along the *c*-axis. The top right-inset plots the magnetic susceptibility ($\chi$) of the crystal, which was also measured along the *c*-axis in a magnetic field of 1 Tesla in the zero-field-cooled (ZFC) and field-cooled (FC) runs. The top left-inset shows the room-temperature x-ray diffraction profile of the (004) Bragg peak of a single crystal of $SrFeO_{3-\delta}$. The sharpness of the peak (full width at half maximum~ $0.12^0$) indicates the high-quality of the single crystal.[27] Resistivity and magnetization measurements presented in figure 2 are consistent with those reported by Lebon *et al.* for $SrFeO_{3-\delta}$ ($\delta$= 0.19) crystal and therefore reveals that the oxygen content of the single crystal sample is close to that of $SrFeO_{2.81}$.[1,2] A sharp rise in the susceptibility close to the temperature for the susceptibility maximum ($T_m$~ 78 K, top right-inset) indicates a transition from the paramagnetic phase to the antiferromagnetic phase and is consistent with earlier studies.[1,2,8] The ZFC and FC curves exhibit thermal hysteresis, revealing the first-order nature of the phase transition. Thermal hysteresis (thermal hysteresis region, $\Delta T$~ 20 K) in the temperature-dependence of susceptibility in $SrFeO_{2.81}$ has been attributed to the coexistence of paramagnetic and antiferromagnetic phases.[5,6] Additionally, a peak that is characteristic of the residual cubic phase has also been observed at 130 K (indicated by the bar herein) in the temperature-dependence of $\chi$.[1,2,8] However, RXS has recently been performed by co-authors of this work, and their results suggested that the transition at ~130 K is related to the weak localization of the charge on $Fe^{3.5+}$, which is also magnetically active and leads to an anomaly in the paramagnetic region of susceptibility.[13] It is evident from figure 2 that during the warming and cooling processes resistivity exhibits thermal hysteresis both in the *ab*-plane and along the *c*-axis. At room temperature (paramagnetic phase) and ~10 K (antiferromagnetic phase), $SrFeO_{2.81}$ has a resistivity of ~4 and ~$10^4$ mΩ-cm, respectively, reflecting typical metallic and semiconducting behaviors. According to the crystal structure in figure 1(b), at room temperature, $SrFeO_{2.81}$ has three Fe sites with two valence states $Fe^{4+}$ and $Fe^{3.5+}$. The low resistivity in the paramagnetic phase is attributed to the delocalized electronic system with a high electron density of the $Fe^{3.5+}$ state and the low electron density in the $Fe^{4+}$ state, as determined by Mössbauer spectroscopy and refined room temperature neutron powder diffraction.[1,2,12] Moreover, the resistivity of the $SrFeO_{2.81}$ crystal, which is independent of the warming or cooling of the sample and the measurement path, begins to increase as the temperature falls below ~130 K and is consistent with the anomaly in the magnetization data in figure 2. The change in resistivity as the temperature declines is reportedly associated with the beginning of the CO transition and similar to that associated with a MI phase transition[13]. Another signature of the phase transition is observed below ~78 K. This transition is accompanied by a thermal hysteresis that is characteristic of a first-order phase transition. According to the phenomenological theory of phase transition across a first-order MI phase transition, as the temperature is reduced, the volume fraction of the insulating phase increases but the metallic phase does not disappear (According to the results of Mössbauer spectroscopy, the calculated fraction of $Fe^{3.5+}$ in the paramagnetic/metallic phase is ~55%, decreasing to ~10% in the antiferromagnetic/insulating phase. Lebon et al.[1] found that the residual ~10% is caused by an oxygen deficiency), which results in the coexistence of two electronic phases over a certain temperature range, giving rise to the hysteresis of electronic properties.[31] A recent study by Lee et al.[13] on $SrFeO_{2.81}$ with majority tetragonal phase revealed that magnetic and charge-related degrees of freedom are coupled with each other and thermal hysteresis in the resistivity is associated with the commensurate-to-incommensurate CO transition (The delocalized $Fe^{3.5+}$ state with fractional valence changes to localized $Fe^{3+}$ and $Fe^{4+}$ states upon the CO transition). Therefore, the paramagnetic-to-antiferromagnetic transition in $SrFeO_{2.81}$ coincides with a commensurate-to-incommensurate CO transition.[13] As expected, the resistivity of a single crystal of $SrFeO_{2.81}$, plotted in figure 2, in the *ab*-plane is always larger than that along the *c*-axis, because the *a* and *b* lattice parameters, as reported by Reehuis et al.,[14] are greater than *c* at all temperatures, possibly lessening the conductivity of electrons. However, the results herein reveal another interesting phenomenon, which is a large difference between the thermal hysteresis region in the *ab*-plane (thermal hysteresis region, $\Delta T \sim 19$ K) and that along the *c*-axis ($\Delta T \sim 30$ K). The anisotropy or direction-dependence of resistivity in the thermal hysteresis region of a single crystal of $SrFeO_{2.81}$ is not mostly explained by the coexistence of antiferromagnetic and paramagnetic phases, by the incommensurate-to-commensurate CO transition,[5,6,13] or in terms of the charge-disproportionation or CDW state.[3,19] Several theoretical and experimental studies of single crystals and thin films of various materials have established that the anisotropy of resistivity is associated with OO or with the preferential occupation of orbitals, is primarily controlled by doping or lattice distortion/strains.[21,23,26,29-33] Accordingly, the anisotropy of resistivity both in the *ab*-plane and along the *c*-axis in $SrFeO_{2.81}$ crystal in the thermal hysteresis region is closely related to the changes in electronic and lattice structures.

Figures 3(a)-(d) present the temperature-dependent Fe *K*-edge XANES spectra of single crystalline $SrFeO_{2.81}$, at two angles of incidence, $\theta = 0°$ and $70°$, to the normal of the *ab*-plane during warming and cooling processes. Corresponding spectra of FeO, $Fe_3O_4$, and $Fe_2O_3$ powder samples were obtained at room temperature, $\theta = 0°$, for reference. According to the dipole-transition selection law, these Fe *K*-edge XANES

spectra are primarily associated with the Fe $1s \rightarrow 4p$ transition, and the intensity of the main feature is attributable to the density of the unoccupied Fe $4p$ states. The bottom panels in figures 3(a)-(d) present the derivative results, to elucidate the variation of the valence state of the Fe ion in $SrFeO_{2.81}$ with temperature, based on the position of the threshold feature.[33] The energy threshold of the Fe $K$-edge feature of $SrFeO_{2.81}$ (feature **d**) is at 7127.5± 0.3 eV for both $\theta= 0°$ and 70° in warming and cooling at all temperatures; this threshold is above those of $FeO/Fe^{2+}$ (feature **a**), $Fe_3O_4/Fe^{(8/3)+}$ (feature **b**), and $Fe_2O_3/Fe^{3+}$ (feature **c**). From the maximum of the derivative of the XANES spectrum, Blasco et al.[3] obtained an average valence state of +4 for Fe ions in the powder sample of $SrFeO_{2.96}$ with an energy threshold at 7127.5± 0.2 eV, which is close to that of $SrFeO_{2.81}$ as displayed in figure 3. However, in the absence of standard Fe $K$-edge XANES for the 4+ valence state, the average valence state of Fe in $SrFeO_{2.81}$ is determined to be between $Fe^{3+}$ and $Fe^{4+}$ and close to $Fe^{4+}$. Additionally, the observed average valence state of Fe in $SrFeO_{2.81}$, as stated above, obtained on the basis of Mössbauer spectroscopic results and the bond valence calculations, is determined to be $Fe^{4+}$, $Fe^{3.5+}$ and $Fe^{4+}$ at the three crystallographic sites, in the ratio 1:2:1.[1,2,12] Importantly, the MI/CO transition of $SrFeO_{2.83\pm0.01}$ and $La_{1/3}Sr_{2/3}FeO_3$ has been attributed to charge-disproportionation of $Fe^{3.5+}$ and $Fe^{3.66+}$, respectively, and new models, based on a CDW,[19] have been proposed to explain the charge modulation in these compounds.[13,19] However, as shown in figures 3(a)-(d), the energy of Fe $K$-edge and the line-shape of SrFeO$_{2.81}$ do not vary much with temperature during warming or cooling. Clearly, the Fe $K$-edge XANES studies do not support the claim that charge disproportionation influences resistivity (or the MI/metal-to-semiconductor transition) below the transition temperature of SrFeO$_{2.81}$.

Figures 4(a) and (b) present the temperature-dependent Fe $L_{3,2}$-edge XANES and XLD (bottom) spectra with **E** parallel to the **ab**-plane (angle of incidence, $\theta = 0°$) and the **c**-axis (angle of incidence, $\theta = 70°$), respectively. The Fe $L_{3,2}$-edge XANES spectra include two features- an $L_3$-edge around 708 eV and an $L_2$-edge around 720 eV, that are separated by spin-orbital splitting. These spectra are primarily associated with the Fe $2p \rightarrow 3d$ transition, and the intensity of the main feature is attributable to the density of the unoccupied Fe $3d$-O $2p$ hybridized states. Notably, as displayed in figures 4(a) and (b), the intensity of the Fe $L_{3,2}$-edge XANES spectra is greatly suppressed when **E** is parallel to the **ab**-plane and almost parallel to the **c**-axis at 60 K during warming and cooling. The Mn$^{3+}$ ion ($d^4$) in an octahedral oxygen environment is well known to exhibit a large static Jahn-Teller (JT) distortion, which has an important role in the CO and MI transition.[20,21,34] SrFeO$_{2.81}$ has two valence states Fe$^{3.5+}$ and Fe$^{4+}$, of which Fe$^{4+}$ ($d^4$) is isoelectronic with Mn$^{3+}$ and is therefore expected to exhibit a JT effect.[34,35] Bocquet et al.[36] noted that $d^4$ is a high-spin state with three electrons' filling: the $t_{2g}$ orbital and the remaining $e_g$ electron itinerant, so $Fe^{4+}$ may induce a strong JT distortion and further split the 3d $e_g$ state into $3d_{x^2-y^2}$ and $3d_{3z^2-r^2}$.[35,36] Figure S1 in the Supplementary Information schematically depicts the experiment to elucidate the in-plane and out-of-plane Fe 3d-O 2p hybridization states. The hybridizations of Fe $3d_{x^2-y^2}$-O $2p_{x,y}$ (in-plane, $\theta= 0°$) and Fe $3d_{3z^2-r^2}$-O $2p_z$ (out-of-plane, $\theta= 70°$) are probed with the electric field $E$ parallel to the $ab$-plane and almost parallel to the $c$-axis, respectively. The XLD spectra (difference in XANES intensity between $\theta= 0°$ and $\theta= 70°$) of the Fe $L_{3,2}$-edge provide information on the preferred orbital of the Fe 3d electron, which is determined by the anisotropic effect with temperatures. The bottom panel in figure 4(b) reveals that the sign of the XLD feature is negative during all cooling, suggesting that Fe $e_g$ electrons preferentially occupy the out-of-plane $3d_{3z^2-r^2}$ orbitals. However, in the warming process, the signs of the XLD spectra are reversed, being positive at 40 and 60 K, suggesting that the Fe $e_g$ electrons preferentially occupied the in-plane $3d_{x^2-y^2}$ orbitals in the thermal hysteresis. This behavior with respect to the preferential orbital can give rise to the anisotropy that is observed in the resistivity in the thermal hysteresis region, owing to the strong coupling between the lattice distortion and the electronic structures.[15-18] Several researchers have provided evidence of structural distortions of transition-metal oxides and their correlation with OO and CO.[15-18,37,38] The distortion of the unit cell is primarily responsible for the lowering of the energy of either out-of-plane $e_g$ orbitals or in-plane $e_g$ orbitals.[15-18] In single crystals of $Pr_{0.5}Ca_{1.5}MnO_4$, an increase in orthorhombic distortion causes OO and changes the electronic properties.[17] However, in $(La_{2-2x}Sr_{1+2x})Mn_2O_7$, competition between local lattice distortion and the hopping of charge carriers is responsible for the comparable amounts of $3d_{3z^2-r^2}$ and $3d_{x^2-y^2}$ orbitals.[16] Raman, far-infrared ellipsometric and neutron diffraction studies of tetragonal $SrFeO_{3-\delta}$ have revealed a change in lattice structure as the temperature falls below $T_m$.[1,2,14] Local lattice distortion in the $SrFeO_{2.81}$ below $T_m$ is evident in the Fe $K$-edge EXAFS results. Tetragonal $SrFeO_{3-\delta}$ has three Fe sites, in pyramidal, distorted/tilted octahedral and octahedral networks of oxygen around Fe ions. The distortion of the oxygen octahedral/pyramidal networks by a change in the nearest-neighbor (NN) Fe-O bond length and the hybridization of Fe $3d$-O $2p$ can stabilize different out-of-plane and in-plane Fe $3d$ $e_g$ orbitals, as evidenced by the temperature-dependent XLD spectra of $SrFeO_{2.81}$ in figure 4.[15-18] Figures S2(a)-(d) in the Supplementary Information plot the temperature-dependent Fourier transform (FT) Fe $K$-edge EXAFS spectra of $SrFeO_{2.81}$ at $\theta= 0°$ (**E**//***ab***-plane) and $\theta= 70°$ (**E**//***c***-axis). The insets present corresponding EXAFS $k^2\chi$ data. The Fe $K$-edge EXAFS spectra include main FT features (**A**, **B** and **C**) that correspond to the NN bond lengths of Fe-O, Fe-Sr and Fe-Fe, respectively, in the $SrFeO_{2.81}$ crystal.[3,39,40]

To obtain detailed information about the temperature dependence of the local structure around the Fe atoms, figure 5 presents a magnified view of feature **A** (corresponding to the NN Fe-O bond length). The figure reveals that the intensity of FT feature is minimal at 300 K for both ***E*//*ab*-plane** and ***E*//*c*-axis** in both warming and cooling processes. However, in figure S2, the intensities of FT features **B** and **C** (corresponding to the NN Fe-Sr and Fe-Fe bond lengths, respectively) follow the typical thermal trend: as the temperature increases, the intensity of each FT feature decreases, suggesting the importance of Fe-O bond lengths in the local distortions of the $SrFeO_{2.81}$ crystal. Evidently, as the temperature is increased from 40 K to 80 K, the intensity of the FT feature increases, before decreasing with a further increasing in temperature. These changes are marked in the warming and cooling cycles for the ***E*//*ab*-plane** but also exist for ***E*//*c*-axis**. The intensity of the FT feature is determined by the coordination number, N and the DW factor $\sigma^2$.[41,42] Figures S2(a)-(d) and figures 5(a)-(d) plot the EXAFS measurements in the ***ab*-plane** (***E*//*ab*-plane**, $\theta = 0°$) and along the ***c*-axis** (***E*//*c*-axis**, $\theta = 70°$): Normally, N does not vary with temperature, so a change in $\sigma^2$ causes the intensity of the FT features to vary with temperature. The DW factor $\sigma^2$ is composed of two components, $\sigma^2_{stat}$ and $\sigma^2(T)_{vib}$, which are associated with static disorder and thermal vibrations, respectively.[42] Component $\sigma^2_{stat}$ is related to the atomic structure and is unrelated to temperature, while the $\sigma^2(T)_{vib}$ is associated with the lattice vibrations, which commonly become smaller as the temperature decreases, according to the Einstein or Debye model.[28,41,42] As expected, at high temperature (above $T_m$), reducing the temperature increases the intensity of the FT feature of the metallic/paramagnetic phase in the SrFeO$_{2.81}$ crystal, owing to the key $\sigma^2(T)_{vib}$ factor, but below $T_m$, the intensity of the FT feature decreases as the temperature declines. These anomalous results clearly reveal that $\sigma^2_{stat}$ dominates the FT intensity below $T_m$, suggesting that static disorders or JT distortions that are caused by Fe 3*d* electrons have a stronger effect than the influence of temperature. The large static distortions of the octahedral/pyramidal oxygen network around the Fe ions below $T_m$ can be understood as being statically contributed to static DW factors that strongly influence the FT feature of Fe-O bonds in the SrFeO$_{2.81}$.

To discuss quantitatively the local atomic structure of a single crystal of SrFeO$_{2.81}$, fitting results for the NN Fe-O bond length, N and the DW factors are obtained using Artemis software[41,42] and presented in Table I of the Supplementary Information and figure 6. The DW factors and NN Fe-O bond lengths are plotted in figure 6 and exhibit anisotropic behavior in the *ab*-plane and along the *c*-axis. Generally, the DW factors that are related to atomic structure should decrease upon cooling of the sample as the thermal vibrations become less intense. However, DW factors in the *ab*-plane abruptly increase when the temperature declines below $T_m$. This sudden increase below the transition temperature [figure 6(a)] with thermal hysteresis indicates that the crystal structure of $SrFeO_{2.81}$ exhibits much greater static disorder below $T_m$ than above $T_m$, and this phenomenon of large static disorder dominates the thermal effect. This fact can be understood with reference to soft phonon mode behavior,[43,44] which is related to a decrease or breaking of the crystal symmetry in the *ab*-plane of $SrFeO_{2.81}$. Soft phonons are typically associated with phase transitions in crystals that can have more than one distinguishable lattice. The *off*-centering of the Fe ions or an order-disorder phase transition between dynamic and static distortions of Fe ions, in which interaction (that is phonon-mediated) between *off*-centered Fe and oxygen atoms drives phonon softening in the $FeO_6$/$FeO_5$ networks. Also, as shown in figure 6(a), DW factors along the *c*-axis are nearly constant as the temperature is reduced, indicating that the thermal effect compete with static disorder along the *c*-axis. The unusually high DW factors in the *ab*-plane and the thermal hysteresis suggest that the local structural ordering of $SrFeO_{2.81}$ differs between the *ab*-plane and the *c*-axis in both warming and cooling processes. The significant changes of DW factors in the *ab*-plane in comparison to *c*-axis are consistent with Reehuis *et al.*[14] In ref. 14, using neutron diffraction studies prominent changes in the bond lengths in *ab*-plane below transition temperature have been reported and it is anticipated to be due to orbital polarization. However, the greater value of DW factors in the *ab*-plane during cooling than warming below $T_m$ is not currently explainable. We speculate that this difference may be related to the difference between the out-of-plane and in-plane Fe 3$d$ orbitals, as observed in the temperature-dependent XLD spectra in figure 4. Therefore, a detailed theoretical calculation may provide a quantitative description of competing lattice, orbital and spin-related degrees of freedom in this system. Figure 6(b) presents the variation of the NN Fe-O bond lengths in the *ab*-plane and along the *c*-axis. As evidenced from figure 6(b), the NN Fe-O bond lengths exhibit hysteretic behavior below the transition temperature $T_m$; increase from 1.91±0.01 to 1.93±0.01 Å in the *ab*-plane, and increase from 1.91±0.01 and 1.92±0.01 Å along *c*-axis. Although the NN Fe-O bond lengths depend differently on temperature in the *ab*-plane and along the *c*-axis, the limitation on the resolution of the EXAFS analysis prevents precise determination of the Fe-O bond lengths. However, the anomalous variations of DW factors in the *ab*-plane and along the *c*-axis suggest that anisotropic distortion of the $FeO_6$/$FeO_5$ network and a change in XLD during warming and cooling processes are responsible for the anisotropy of the resistivity in the thermal hysteresis region. These results further demonstrate that the instabilities in the local Fe-O bond length/DW factors and preferred Fe 3$d$ $e_g$ orbitals drive the MI/metal-to-semiconductor transition in $SrFeO_{2.81}$, in a manner similar to the driving of the Peierls MI transition in $VO_2$, which was elucidated by Budai *et al.* from first-principle calculations.[44] Their calculations also revealed that increased occupation of V $3d_{x^2-y^2}$ orbitals induces the Peierls instability; lowers the total electronic energy, and opens the insulating band gap.[44] As discussed above, the XLD studies herein of the SrFeO$_{2.81}$ crystal reveal similar phenomena. During the warming process below the transition temperature, the preferential occupation changes to in-plane Fe $3d_{x^2-y^2}$ orbitals and these changes are associated with anisotropic distortion of the FeO$_6$/FeO$_5$ network during both warming and cooling.

The above assertion may also be related to possible CDW behavior in SrFeO$_{2.81}$.[45,46] As stated earlier, Lee et al.[13] used RXS to provide evidence that satellite peaks are induced by charge modulation below the transition temperature in single crystalline tetragonal SrFeO$_{2.81}$. As a consequence of an electron-phonon interaction, a CDW is associated with charge modulation and lattice distortion in SrFeO$_{2.81}$,[13] as in other CDW materials.[3,19,47] Lattice distortion is also evident from the anomalous DW factors/bond lengths below T$_m$ for the single crystal of SrFeO$_{2.81}$, as depicted in figure 6. Furthermore, a CDW is a charge/electronic modulation process that arises from the coupling of valence and conduction bands, opening a band gap at/near the Fermi level ($E_F$).[48,49] Therefore, to further examine the band gap opening across a MI/metal-to-semiconducting transition in SrFeO$_{2.81}$, VB-PES spectra with photon energy hυ= 58 eV and O $K$-edge XANES spectra of SrFeO$_{2.81}$ were obtained during warming and cooling processes, and presented in figures 7(a)-(d) for $E//ab$-plane and $E//c$-axis. The VB-PES spectra exhibit three major features, $b_1$, $b_2$ and $b_3$. The line shape of each feature is similar to those in the experimental spectra of $SrFeO_{3-\delta}$, $LaFeO_3$ and Sr-doped $LaFeO_3$.[50-52] According to the theoretical studies of $SrFeO_{3-\delta}$[50] and Sr-doped $LaFeO_3$[51-53] at low photon energies ($20 \leq h\nu \leq 100$ eV) the O 2$p$ cross-section dominates the Fe 3$d$ emission, so the region (from 0 to ~2 eV) at/below $E_F$ can be attributed to the O 2$p$ and Fe 3$d$ ($e_g$) hybridized states, which are essential to have a critical role in the band gap opening in $SrFeO_{2.81}$, as discussed below. The O $K$-edge XANES spectra of $SrFeO_{2.81}$ reflect transitions from O 1$s$ to unoccupied 2$p$ states. The empty O 2$p$ states are hybridized with the 3$d$ and 4$sp$ bands of Fe with Sr 4$d$ bands,[4] so the features that are indicated by $a_1$, $a_2$, $a_3$ and $a_4$ in figures 7(a)-(d) correspond to the hybridized O 2$p$-Fe 3$d$ states that are subject to crystal-field splitting, forming $t_{2g\downarrow}$, $e_{g\uparrow}$, $e_{g\downarrow}$ and $e_{g\downarrow}$ states, respectively, above/near the $E_F$, as determined from earlier theoretical work.[4] The insets in figures 7(a)-(d) magnify VB-PES and O $K$-edge XANES spectra near $E_F$, to elucidate the *relative band gap* opening (or energy separation) as a function of temperature. To calculate the band gap, the leading edges in both the VB-PES and O$K$-edge XANES spectra of $SrFeO_{2.81}$ are extrapolated to the baselines to obtain valence-band maximum ($E_{VBM}$) and conduction-band-minimum ($E_{CBM}$),[51,54] respectively. At room temperature, the two extrapolated lines intersect each other, indicating the SrFeO$_{2.81}$ sample has no band gap and exhibits metallic behavior, which is consistent with the resistivity measurement. The fact that the value of band gap at room temperature is zero in both directions, however, during both warming and cooling processes a semiconducting band gap opens up at/near the $E_F$ and the relative value of the band gap increases as the temperature decreases. At 40 K, these values are ~0.8 and 0.6 eV for ***E*//*ab*-**plane and ~0.7 and 0.6 eV for ***E*//*c*-**axis, during warming and cooling processes, respectively. These results also reveal that the band gaps are almost independent of direction and whether the process in question is warming or cooling. The above results provide further evidence of an abrupt increase in resistivity at ~130 K. However, CDW modulations and the relative band gaps do not exhibit the anisotropy close to the thermal hysteresis region (~78 K), indicating that they may not depend on direction in SrFeO$_{2.81}$ crystal or they are outside the energy resolution limit of VB-PES (~18 meV), owing to the smallness of the relevant changes. The abrupt increase in the resistivity and relative band gap opening at low temperatures provide clear evidence of the CDW nature of single crystalline SrFeO$_{2.81}$. Since resistivity is determined by the electron-phonon coupling, it depends on the lattice distortion. We therefore believe that the sudden increase of resistivity below ~130 K in both directions is associated with the formation of the CDW, this claim is consistent with the appearance of satellite peaks in the RXS studies of Lee *et al.*[13] However, the anisotropic behavior of resistivity in the thermal hysteresis region (~78 K) arises from anisotropic DW factors/bond lengths, as evident from the EXAFS analysis, and further stabilizes the different in-plane and out-of-plane Fe 3$d$ orbitals that are observed from XLD during warming and cooling processes.

In summary, the local electronic and atomic structures of SrFeO$_{2.81}$ were elucidated using temperature-dependent XANES and VB-PES techniques, to determine the origin of the anisotropy of resistivity in the *ab*-plane and along the *c*-axis in the thermal hysteresis region. The Fe $L_{3,2}$-edge XLD results reveal that, during cooling from room temperature to below T$_m$, the Fe 3$d$ electrons preferentially occupy the out-of-plane Fe $3d_{3z^2-r^2}$ orbitals. However, during warming below T$_m$, they preferentially occupy the in-plane Fe $3d_{x^2-y^2}$ orbitals. Local atomic structural analysis of the temperature-dependent Fe $K$-edge EXAFS of SrFeO$_{2.81}$ reveals unusually large DW factors with thermal hysteresis below T$_m$. Additionally, NN Fe-O bond lengths exhibit anisotropy in the *ab*-plane. Experimental results suggest that the local atomic structural ordering in the *ab*-plane differs from that along the *c*-axis during both warming and cooling processes. This distinction stabilizes the difference between the forms of atomic structural ordering in the warming and cooling processes and is responsible for the anisotropy of resistivity below T$_m$ in the thermal hysteresis region in the *ab*-plane and along the *c*-axis in a single crystal of $SrFeO_{2.81}$. The abrupt increase in resistivity and evidence of relative band gap opening that were obtained in the O *K*-edge XANES and VB-PES experiment, along with the presence of satellite peaks in RXS spectra[13], confirm the CDW nature of the $SrFeO_{2.81}$ single crystal at low temperatures.

§ These authors contributed equally to this work.

[†] Present affiliation: Centre of Material Sciences, Institute of Interdisciplinary Studies, University of Allahabad, Allahabad-211002, Uttar Pradesh, India

Correspondence and request for materials should be addresses to wfpong@mail.tku.edu.tw

## Methods

**Sample preparation and characterization:** High-quality single crystals of $SrFeO_{2.81}$ were prepared using the floating zone method.[27] In ref. 13 reported by Lee *et al.*, the estimated number of oxygen-content (2.875) in $SrFeO_x$ is based from their XRD measurements, however, they did not employ techniques such as thermogravimetric analysis of iodometric titration to determine the oxygen content. Meanwhile, the refined parameter of XRD cannot tell structure different between $SrFeO_{2.81}$ and $SrFeO_{2.875}$ because of the insensitivity of X-ray to oxygen atom. Further, according to the refs. 1 and 2, and introduction section of the present work, $SrFeO_x$ with oxygen-content x= 2.875 is reported to be pure tetragonal, however for x= 2.75, it is pure orthorhombic, so our sample (SrFeO$_{2.81}$) with oxygen-content x= 2.81 lies in between these two and also XRD presented in figure 1 confirm the tetragonal nature of the sample. The magnetic susceptibility and electrical resistivity were measured using a superconducting quantum interference device and a physical property measurement system, respectively. XRPD patterns were obtained on an image plate at beamline-01C of the National Synchrotron Radiation Research Center (NSRRC) in Hsinchu, Taiwan, using X-ray with a wavelength of 0.728 Å (16 keV). Lebail refinements of XRPD patterns were performed using the Fullprof software package.[55-57] Fe $L_{3,2}$-edge XANES/XLD, $K$-edge XANES/EXAFS, O $K$-edge XANES and VB-PES (incident photon energy, hυ= 58 eV) experiments were conducted at the Dragon-11A, Wiggler-17C and U9-ARPES BL21B1 beamlines at the NSRRC. The measurements were made during warming and cooling at various temperatures to elucidate the local electronic and atomic structures of the SrFeO$_{2.81}$ single crystal. Further, in the warming and cooling process, all the experiments have been carried out in the same conditions. In the warming process, sample is heated from the lowest temperature (30 K) to 40 K. At 40 K, sample temperature has been kept constant for ~30 mins and then data have been collected for two orientations of sample. This is repeated for all temperatures and during cooling process too. The temperature of the sample was controlled using a closed-cycle refrigerator with an accuracy of ±0.1 K. Local atomic structure analysis of EXAFS data was performed using the Artemis program. Artemis combines the multiple-scattering EXAFS computer program FEFF[41] and the nonlinear least-squares-fitting computer program FEFFIT.[42]

**Figure captions**

**Figure 1: (a)** Observed (red dots), calculated (black line) and difference (bottom green line) patterns of SrFeO$_{2.81}$, obtained using Lebail refinement of synchrotron X-ray powder diffraction data. Vertical check marks above difference profiles indicate Bragg reflections. Insets magnify selected pseudocubic reflections, **(b)** crystalline structure of SrFeO$_{2.81}$; valence Fe$^{4+}$ is attributed to Fe(1), Fe(3) and Fe$^{3.5+}$ is attributed to Fe(2), based on previous work,[1] **(c)** FeO$_5$ square pyramidal, **(d)** FeO$_6$ distorted octahedra, and **(e)** FeO$_6$ octahedra.

**Figure 2:** Temperature-dependence of resistivity of a single crystal of SrFeO$_{2.81}$, measured in *ab*-plane and along *c*-axis. Top inset shows temperature-dependence of magnetic susceptibility ($\chi$) measured along *c*-axis in ZFC and FC runs in a magnetic field of 1 Tesla, and bottom inset presents room-temperature x-ray diffraction profile

showing (004) Bragg peak obtained in $\theta$-scan.

**Figure 3(a)-(d):** The temperature-dependent Fe $K$-edge XANES spectra of single crystal SrFeO$_{2.81}$ measured at two different angles of incidence $\theta= 0°$ (with electric field $E$ parallel to the $ab$-plane) and 70° (with electric field $E$ nearly parallel to the $c$-axis) on warming and cooling process. Corresponding spectra were obtained for FeO, Fe$_3$O$_4$, and Fe$_2$O$_3$ powder samples at room temperature with angle $\theta= 0°$ for reference.

**Figure 4(a) and (b):** Temperature-dependence of normalized Fe $L_{3,2}$-edge XANES spectra of single crystal of SrFeO$_{2.81}$ at two angles of incidence $\theta= 0°$ and 70° during warming and cooling. Bottom panels show corresponding XLD spectra.

**Figure 5:** Temperature-dependence of the main FT feature **A** (corresponding to the NN Fe-O bond distance) of Fe $K$-edge EXAFS for **(a, b)** $E$//$ab$-plane and **(c, d)** $E$//$c$-axis in the warming and cooling process.

**Figure 6:** Variation of **(a)** DW factors and **(b)** NN Fe-O bond lengths with temperature, obtained by fitting temperature-dependent Fe $K$-edge EXAFS for $R$ from 1.15 to 1.96 Å with angle of incidence $\theta=0°$, and $R$ from 1.04 to 1.77 Å with angle of incidence $\theta=70°$.

**Figure 7(a)-(d):** Normalized VB-PES and O $K$-edge XANES spectra of a single crystal SrFeO$_{2.81}$ at two angles of incidence $\theta= 0°$ ($E$//$ab$-plane) and 70° ($E$//$c$-axis) during warming and cooling. VB-PES spectra are obtained at photon energy of 58 eV. Insets display linear fits to VB-PES and O *K*-edge XANES spectra at various temperatures and relative band gaps.

## Acknowledgments

The author (W.F.P.) would like to thank the National Science Council of Taiwan and Ministry of Science and Technology of Taiwan (MoST) for providing financial support for the research under project No. NSC 102-2112-M-032-007-MY3 and NSC 102-2632-M-032-001-MY3. R.S.S. is thankful to the Department of Science and Technology (DST), India for DST INSPIRE Faculty award (DST/INSPIRE/04/2015/002300).

## Author contributions

S.H.H., R.S.S. and W.F.P. designed the experiments having prior discussion with C.H.D. The SrFeO$_{2.81}$ sample was synthesized by C.H.Y., C.H.D. and F.C.C. All measurements are performed by S.H.H., R.S.S., Y.F.W., Y.C.S., S.H.L., H.T.W., J.W.C., Y.Y.C., H.M.T., J.L.C., C.W.P., C.M.C., W.C.C., H.J.L. and J.F.L. The data analysis and manuscript writing are done by S.H.H., R.S.S. and W.F.P. All authors discussed the results and contributed to finalization of the manuscript.

## Additional information

**Competing financial interests**

The authors declare no competing financial interests.

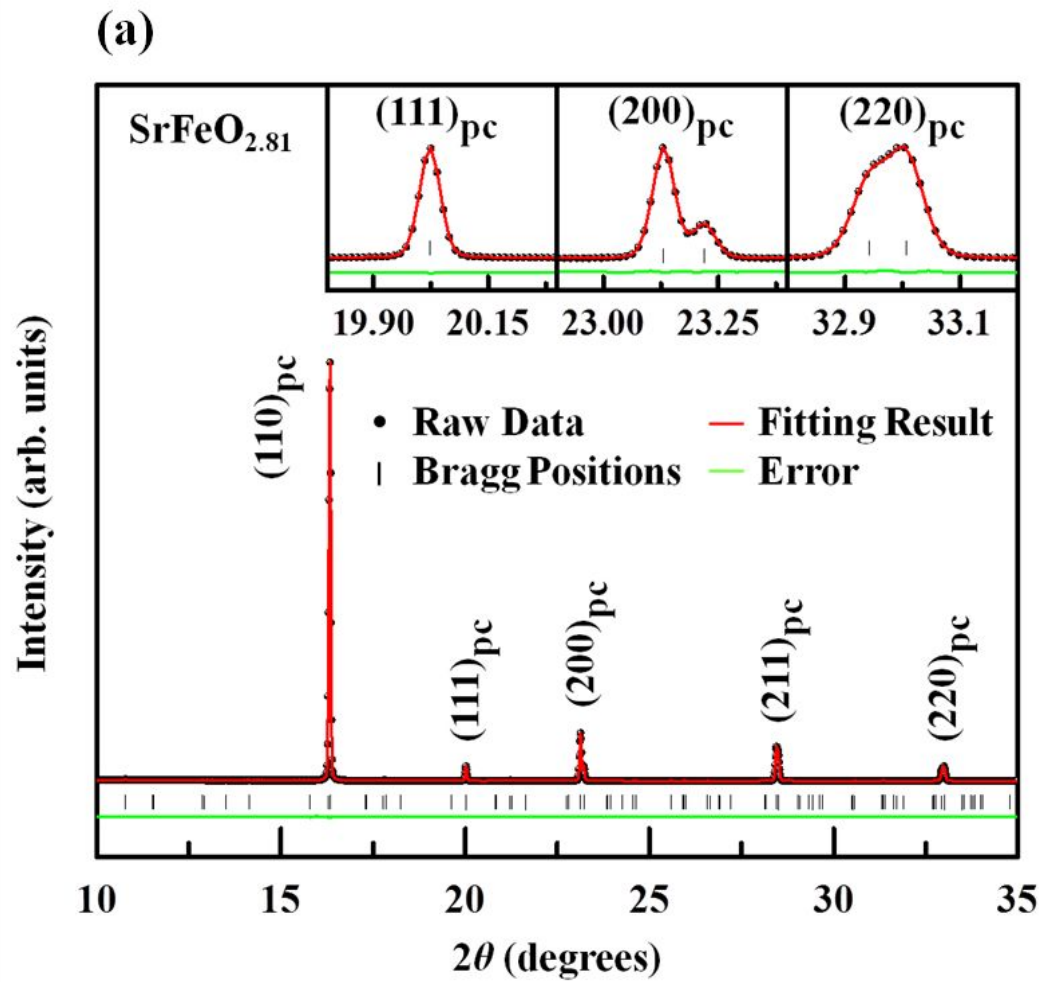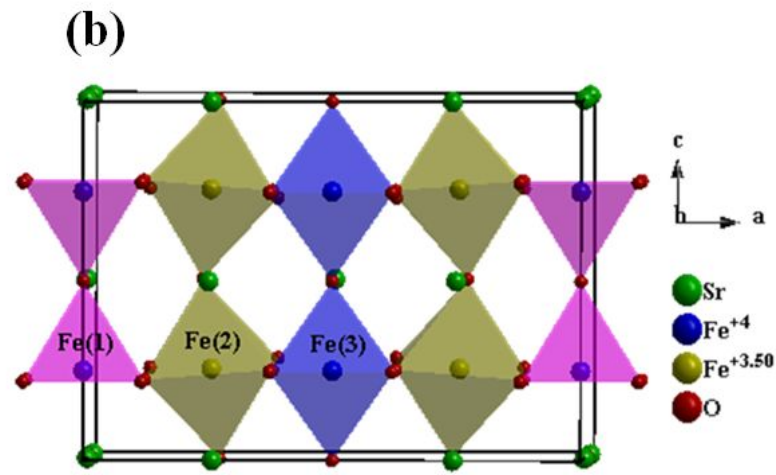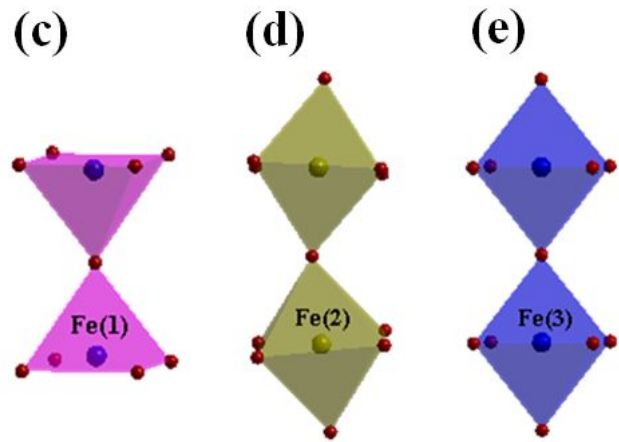

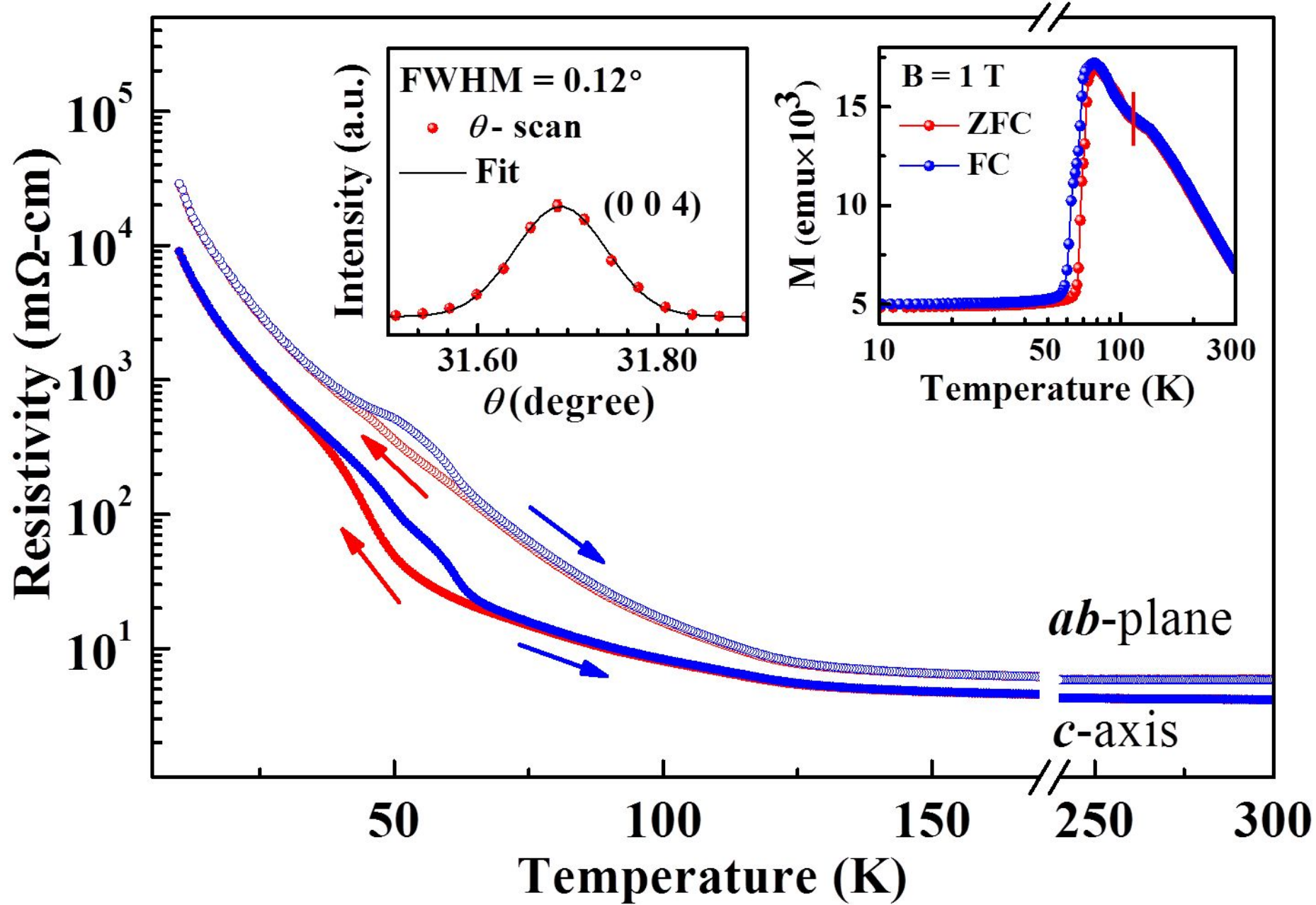

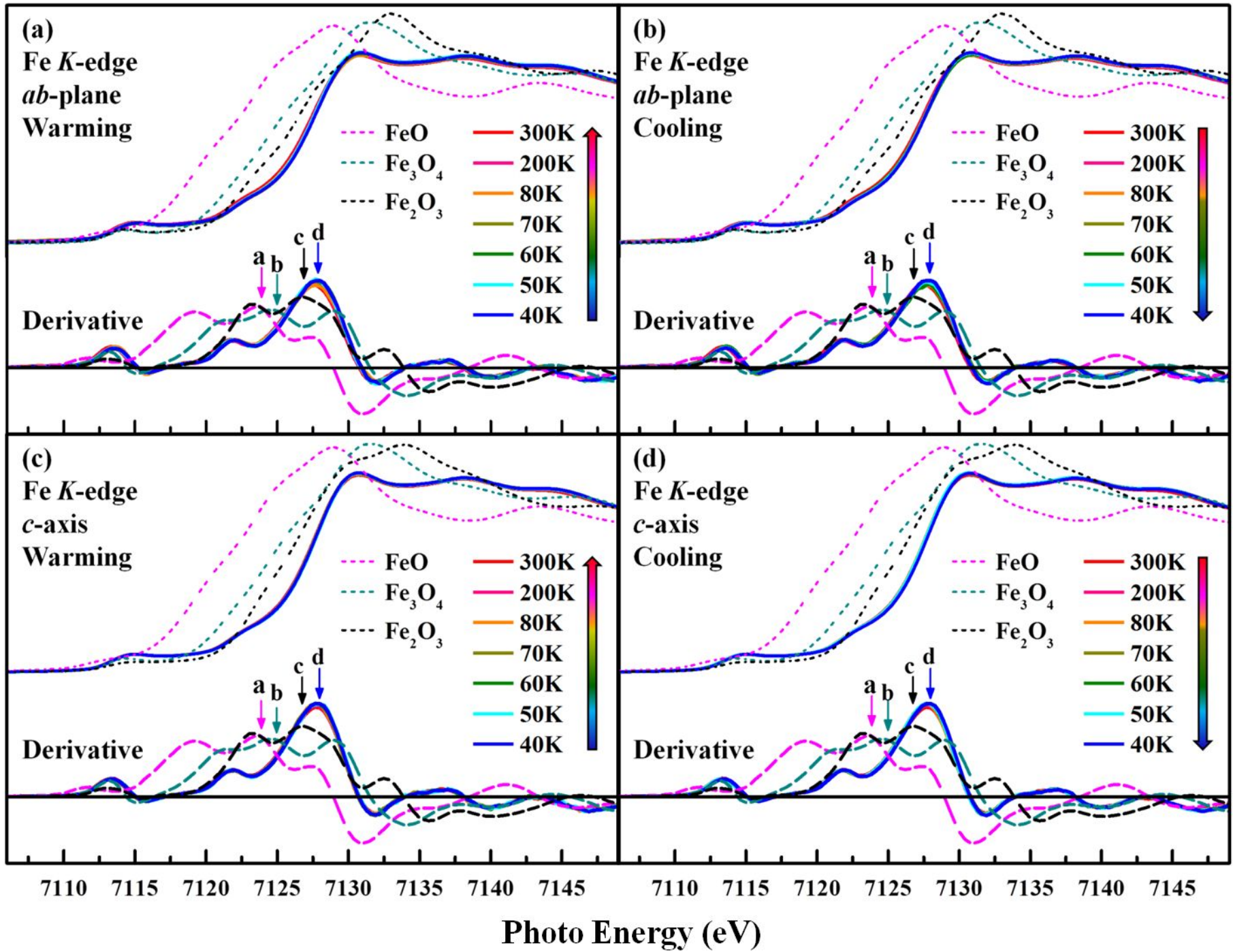

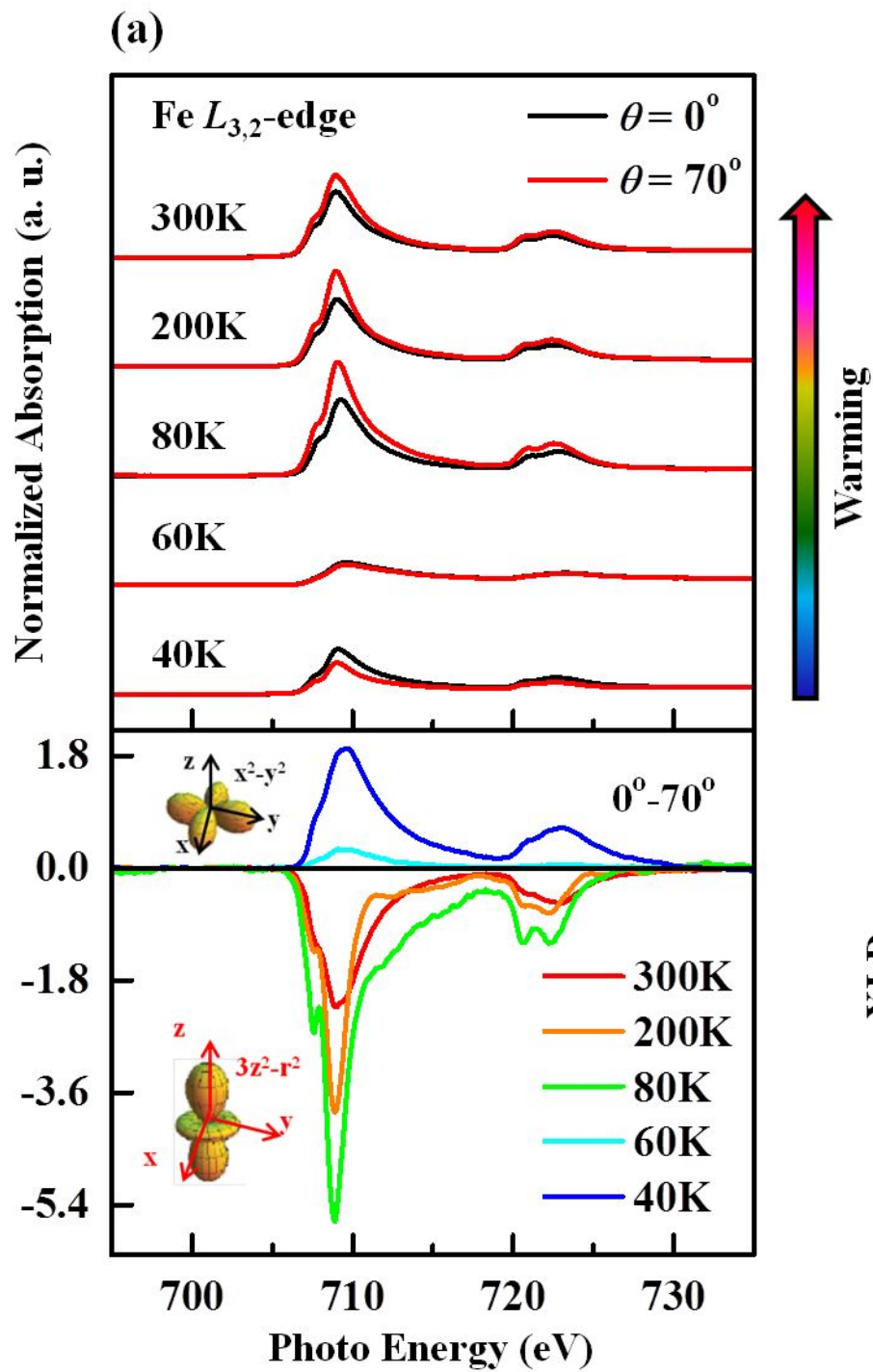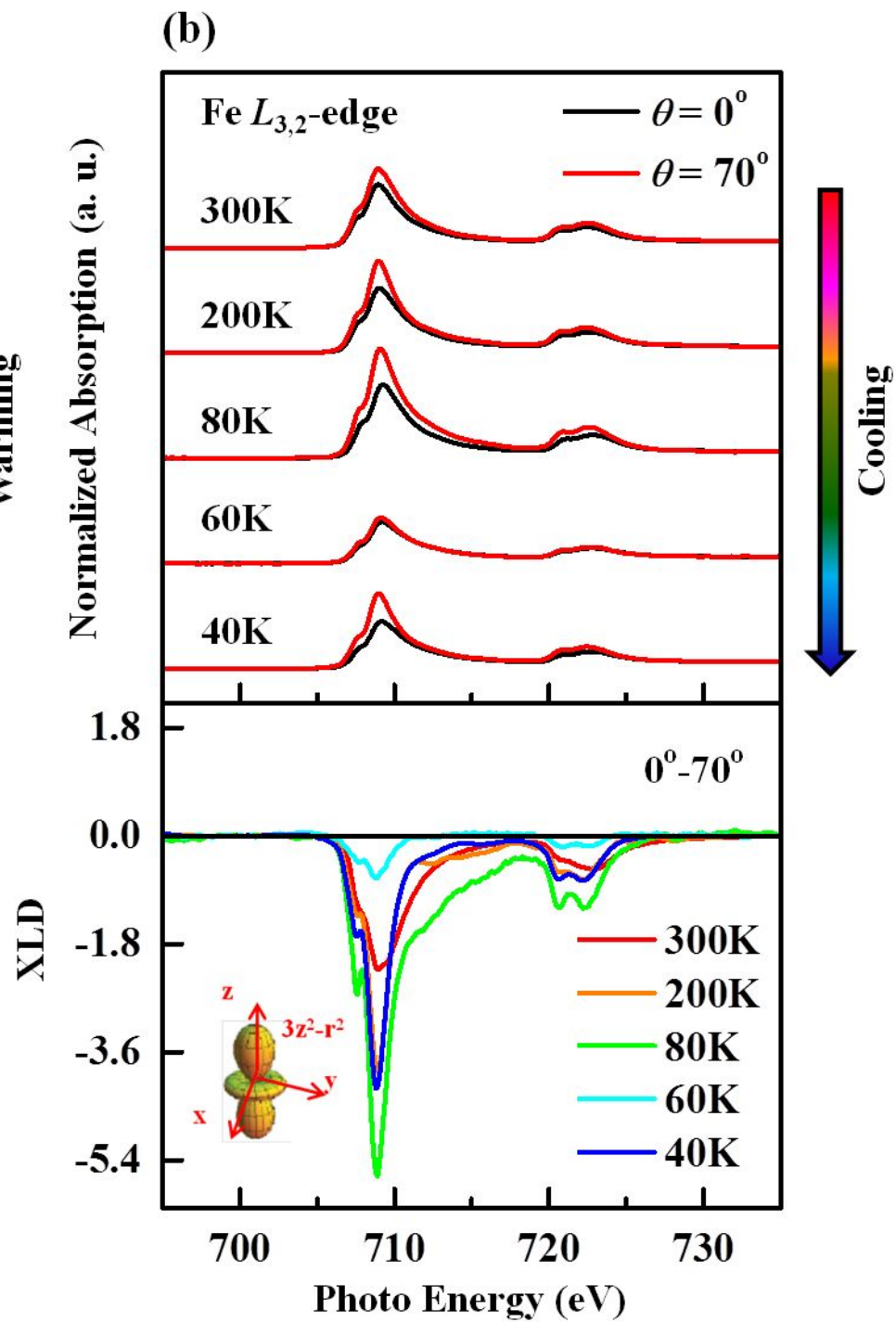

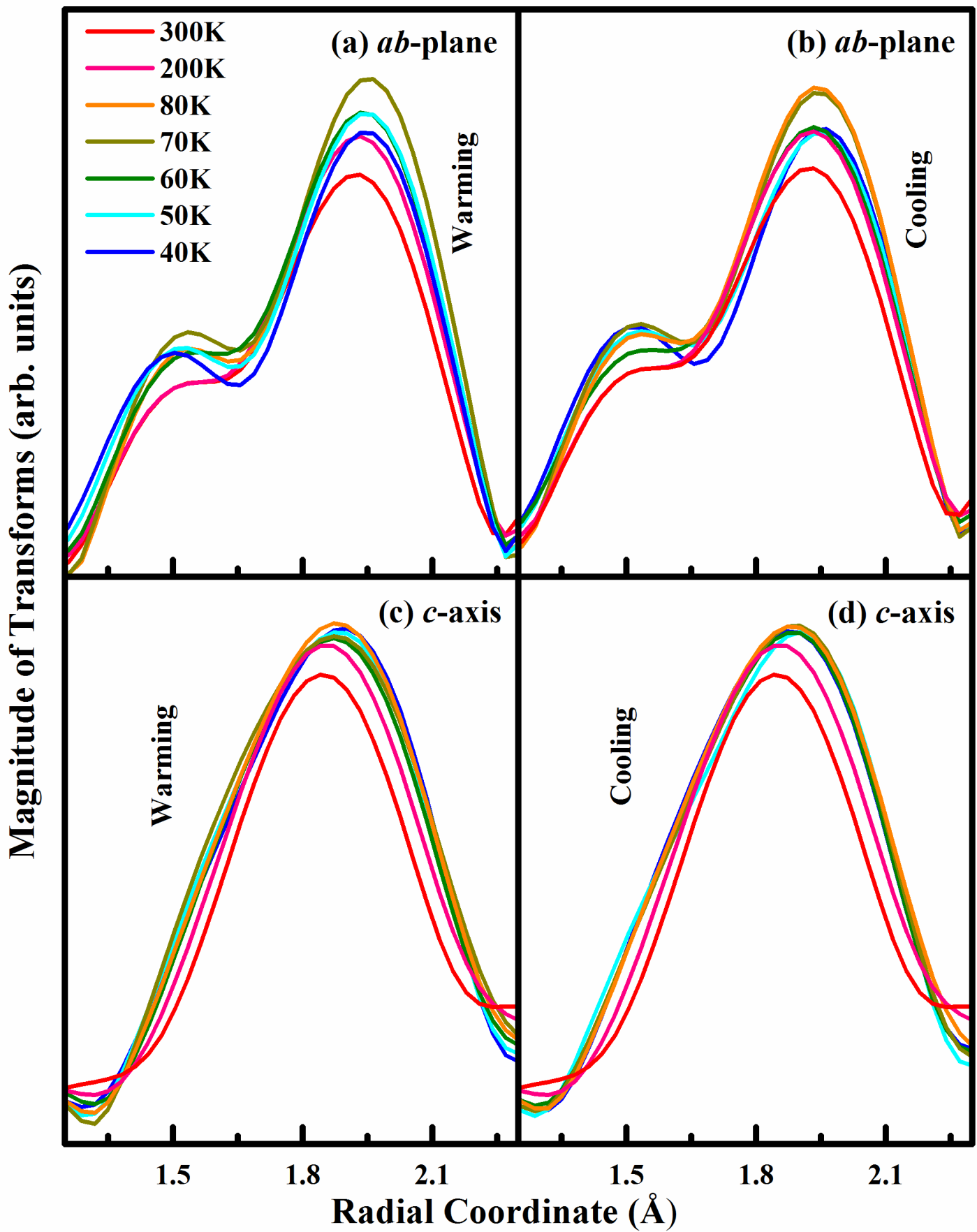

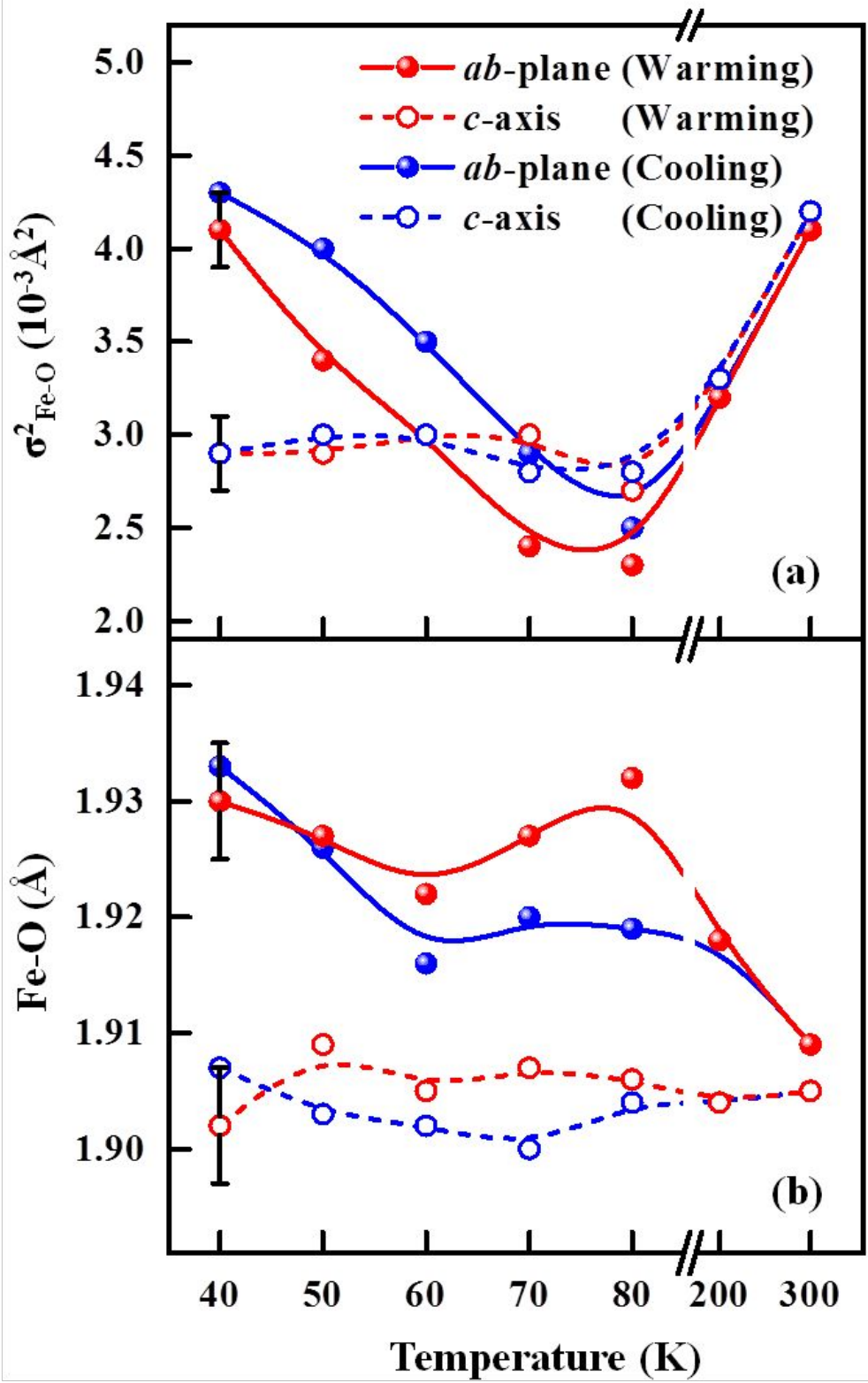

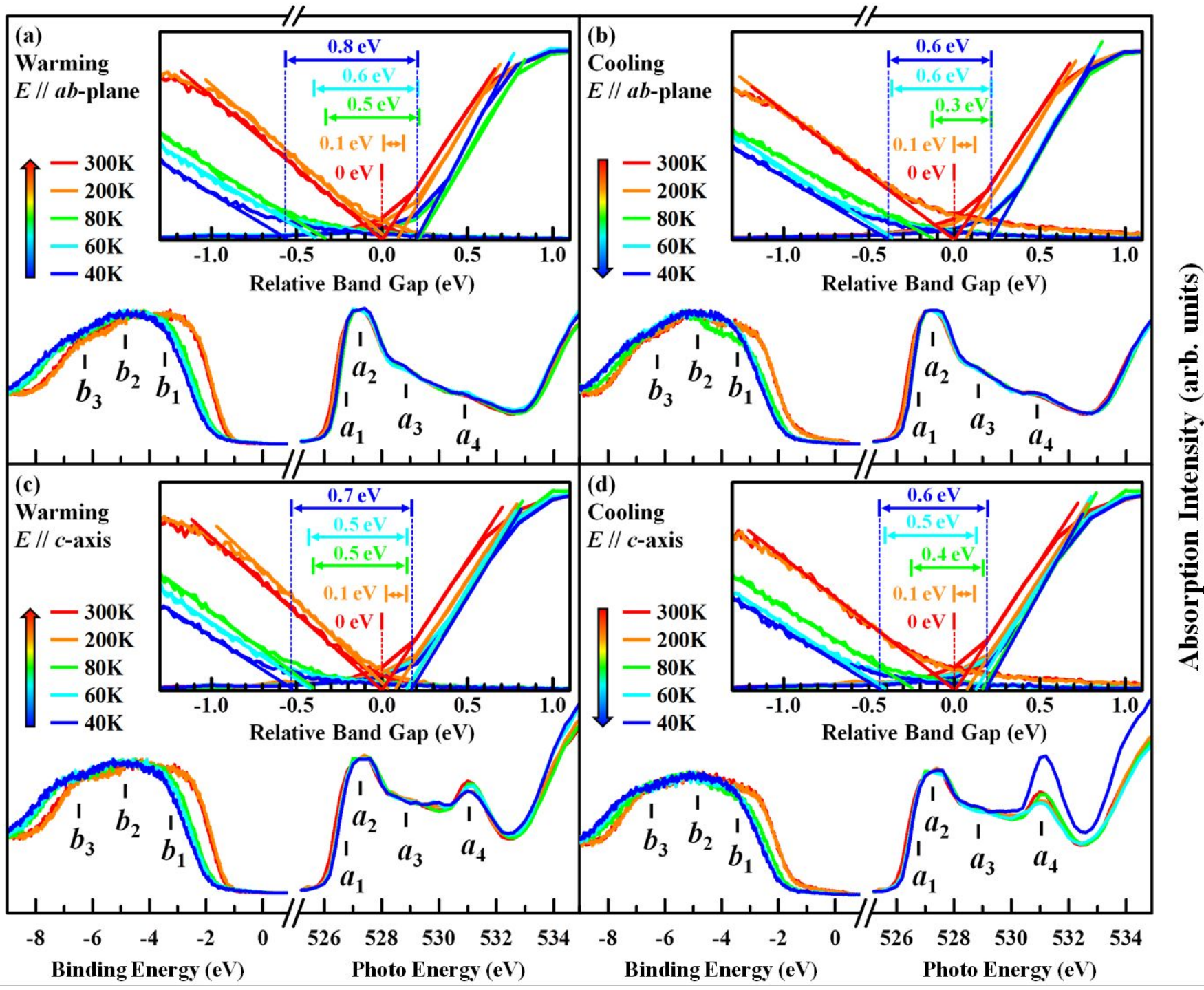

# Supplementary Information

# Anisotropy in the thermal hysteresis of resistivity and charge density wave nature of single crystal SrFeO$_{3-\delta}$: X-ray absorption and photoemission studies

S. H. Hsieh,[1, §] R. S. Solanki,[1, §, †] Y. F. Wang,[1] Y. C. Shao,[1] S. H. Lee,[1] C. H. Yao,[1] C. H. Du,[1] H. T. Wang,[2] J. W. Chiou,[3] Y. Y. Chin,[4] H. M. Tsai,[4] J.-L. Chen,[4] C.W. Pao,[4] C.-M. Cheng,[4] W.-C. Chen,[4] H. J. Lin,[4] J. F. Lee,[4] F. C. Chou,[5] W. F. Pong[1,*]

[1] Department of Physics, Tamkang University, Tamsui 251, Taiwan
[2] Department of Physics, National Tsinghua University, Hsinchu 300, Taiwan
[3] Department of Applied Physics, National University of Kaohsiung, Kaohsiung 811, Taiwan
[4] National Synchrotron Radiation Research Center, Hsinchu 300, Taiwan
[5] Center for Condensed Matter Sciences, National Taiwan University, Taipei 106, Taiwan

**Figure S1:** Schematic representation to elucidate out-of-plane and in-plane 3d states. The direct hybridization of Fe $3d_{x^2-y^2}$-O $2p_{x,y}$ (in-plane) and Fe $3d_{3z^2-r^2}$-O $2p_z$ (out-of-plane) are probed by electric field $E$ parallel to the *ab*-plane (angle of incidence, $\theta = 0$) and electric field $E$ nearly parallel to the *c*-axis (angle of incidence, $\theta = 70$), respectively.

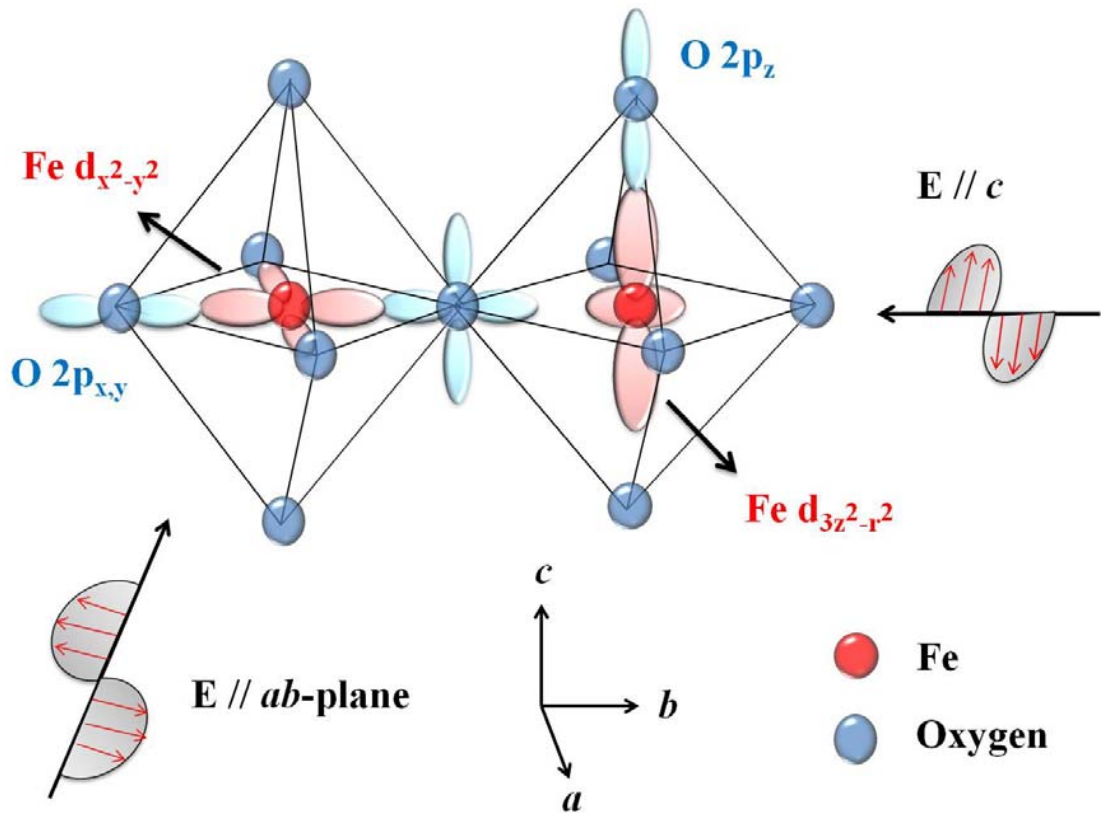

**Figure S2:** The magnitude of the FT spectra of the temperature dependent Fe $K$-edge EXAFS **(a)** and **(b)**: in a $k$ range from 2.65 to 11.51 Å$^{-1}$ at angle of incidence $\theta= 0°$ (**E**//**ab**-plane); **(c)** and **(d)**: in a $k$ range and from 2.65 to 11.51 Å$^{-1}$ at angle of incidence $\theta= 70°$ (**E**//**c**-axis) on warming and cooling process. Three main FT features **A**, **B** and **C**, which correspond to the nearest-neighbor Fe-O, Fe-Sr and Fe-Fe bond distances for SrFeO$_{2.81}$ are marked with vertical lines. Insets show the $k^2\chi$ dependence as a function of $k$.

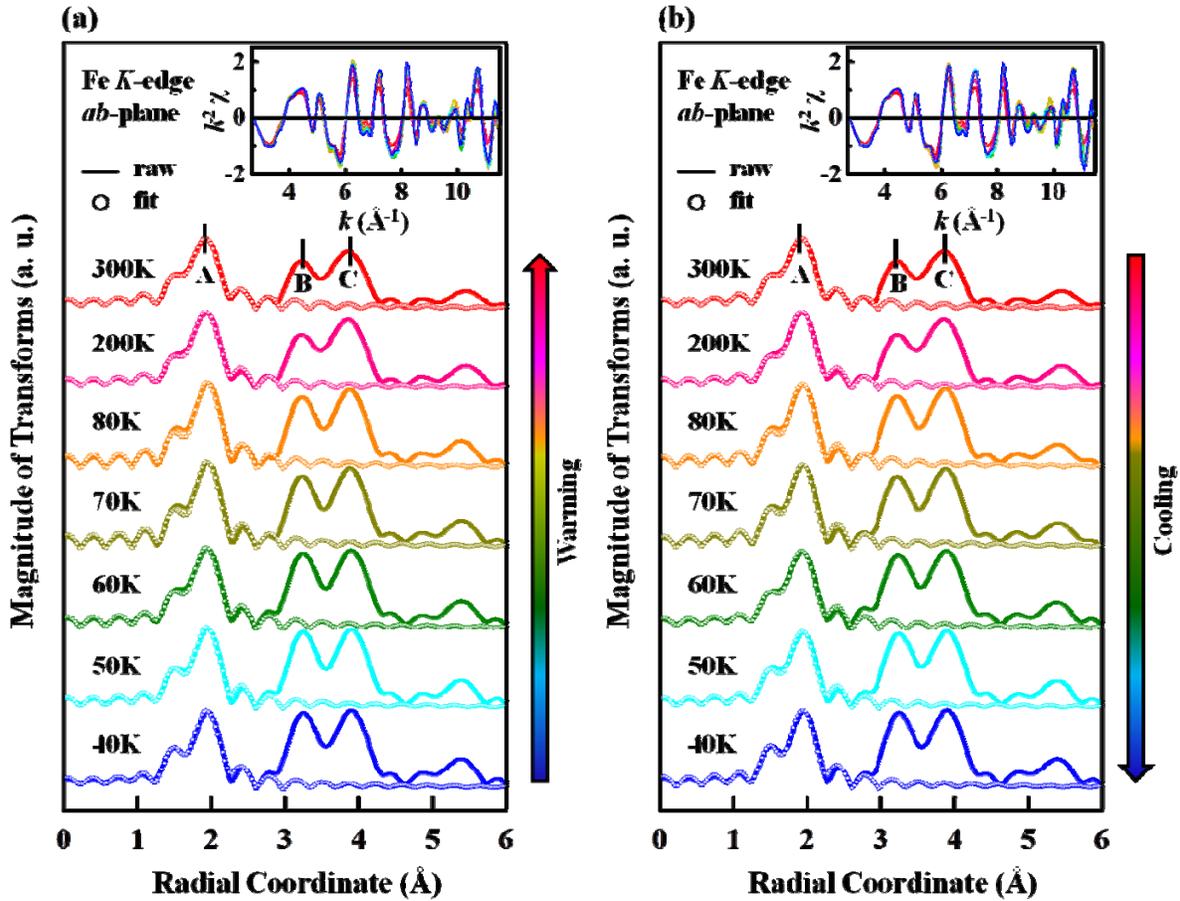

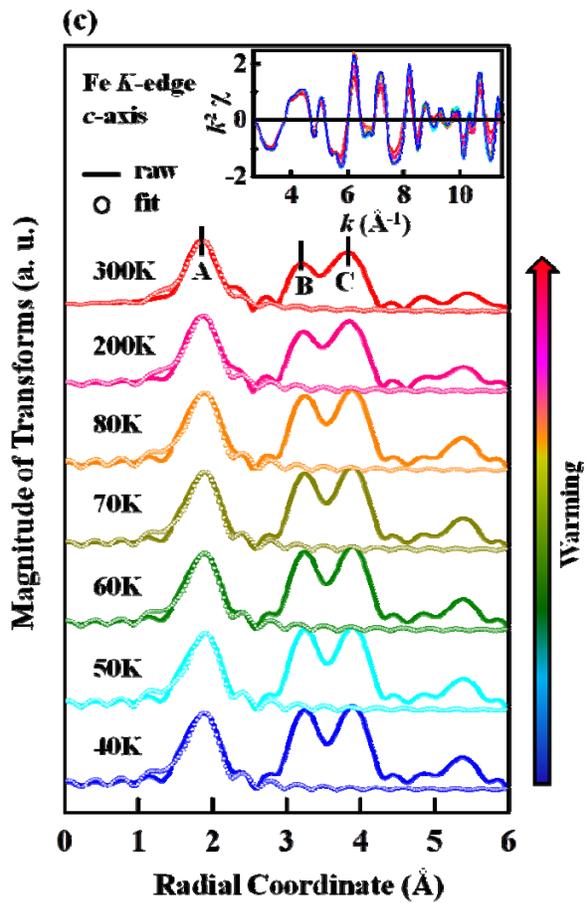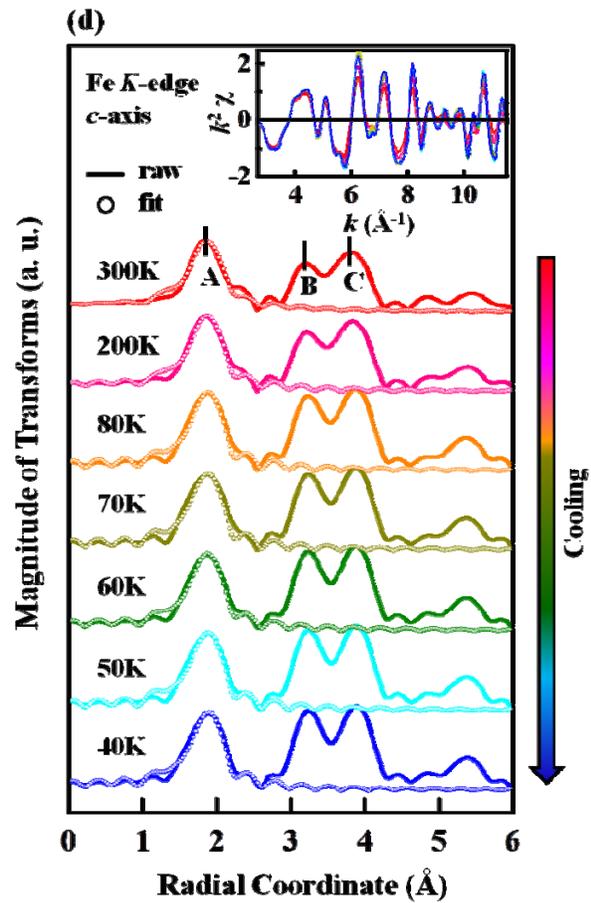

**Table 1:** Best fit parameters obtained from the fitting of Fe $K$-edge EXAFS data in the $R$-space mode from 1.15 to 1.96 Å for angle of incidence $\theta= 0°$ ($E//ab$-plane) and from 1.04 to 1.77 Å for angle of incidence $\theta= 70°$ ($E//c$-axis) during warming and cooling process. $N_{ab}$, $\sigma^2_{ab}$, and $R_{ab}$ & $N_c$, $\sigma^2_{ab}$, and $R_c$ correspond to number of nearest neighbor (NN) oxygen ions around central Fe ion, square of Debye-Waller factor for Fe-O, and average Fe-O bond length in the $ab$-plane and $c$-axis, respectively. In the pyramidal coordination along $c$-axis there is only one NN oxygen for Fe, however in the octahedral they are two, therefore the average value of NN along $c$-axis ($N_c$) has been around 1.5± 0.1, for the $ab$-plane as usual the NN oxygen atoms ($N_{ab}$) for both pyramidal and octahedral coordination are around 4.0± 0.1.

| Temperature | $N_{ab}$ | $N_c$ | $\sigma^2_{ab}$ ($\times 10^{-3}$Å$^2$) | $\sigma^2_c$ ($\times 10^{-3}$Å$^2$) | $R_{ab}$ (Å) | $R_c$ (Å) |
|---|---|---|---|---|---|---|
| **Cooling** | | | | | | |
| 300K | 4± 0.1 | 1.5± 0.1 | 4.1 ± 0.2 | 4.2 ± 0.2 | 1.909 ± 0.005 | 1.905 ± 0.005 |
| 200K | 4± 0.1 | 1.5± 0.1 | 3.2 ± 0.2 | 3.3 ± 0.2 | 1.918 ± 0.005 | 1.904 ± 0.005 |
| 80K | 4± 0.1 | 1.5± 0.1 | 2.5 ± 0.2 | 2.7 ± 0.2 | 1.919 ± 0.005 | 1.904 ± 0.005 |
| 70K | 4± 0.1 | 1.5± 0.1 | 2.9 ± 0.2 | 3.0 ± 0.2 | 1.920 ± 0.005 | 1.900 ± 0.005 |
| 60K | 4± 0.1 | 1.5± 0.1 | 3.5 ± 0.2 | 3.0 ± 0.2 | 1.916 ± 0.005 | 1.902 ± 0.005 |
| 50K | 4± 0.1 | 1.5± 0.1 | 4.0 ± 0.2 | 2.9 ± 0.2 | 1.926 ± 0.005 | 1.903 ± 0.005 |
| 40K | 4± 0.1 | 1.5± 0.1 | 4.3 ± 0.2 | 2.9 ± 0.2 | 1.933 ± 0.005 | 1.907 ± 0.005 |
| **Warming** | | | | | | |
| 40K | 4± 0.1 | 1.5± 0.1 | 4.1 ± 0.2 | 2.9 ± 0.2 | 1.930 ± 0.005 | 1.902 ± 0.005 |
| 50K | 4± 0.1 | 1.5± 0.1 | 3.4 ± 0.2 | 3.0 ± 0.2 | 1.927 ± 0.005 | 1.909 ± 0.005 |
| 60K | 4± 0.1 | 1.5± 0.1 | 3.0 ± 0.2 | 3.0 ± 0.2 | 1.922 ± 0.005 | 1.905 ± 0.005 |
| 70K | 4± 0.1 | 1.5± 0.1 | 2.4 ± 0.2 | 2.8 ± 0.2 | 1.927 ± 0.005 | 1.907 ± 0.005 |
| 80K | 4± 0.1 | 1.5± 0.1 | 2.3 ± 0.2 | 2.8 ± 0.2 | 1.932 ± 0.005 | 1.906 ± 0.005 |
| 200K | 4± 0.1 | 1.5± 0.1 | 3.2 ± 0.2 | 3.3 ± 0.2 | 1.918 ± 0.005 | 1.904 ± 0.005 |
| 300K | 4± 0.1 | 1.5± 0.1 | 4.1 ± 0.2 | 4.2 ± 0.2 | 1.909 ± 0.005 | 1.905 ± 0.005 |